\documentclass{llncs} 

\usepackage{amssymb,stmaryrd}

\newbox\tempa
\newbox\tempb
\newdimen\tempc
\def\mud#1{\hfil $\displaystyle{\mathstrut #1}$\hfil}
\def\rig#1{\hfil $\displaystyle{#1}$}
\def\irulehelp#1#2#3{\setbox\tempa=\hbox{$\displaystyle{\mathstrut #2}$}%
                        \setbox\tempb=\vbox{\halign{##\cr
        \mud{#1}\cr
        \noalign{\vskip\the\lineskip}
        \noalign{\hrule height 0pt}
        \rig{\vbox to 0pt{\vss\hbox to 0pt{${\; #3}$\hss}\vss}}\cr
        \noalign{\hrule}
        \noalign{\vskip\the\lineskip}
        \mud{\copy\tempa}\cr}}
                      \tempc=\wd\tempb
                      \advance\tempc by \wd\tempa
                      \divide\tempc by 2 }
\def\irule#1#2#3{{\irulehelp{#1}{#2}{#3}
                     \hbox to \wd\tempa{\hss \box\tempb \hss}}}
\def\birulehelp#1#2#3{\setbox\tempa=\hbox{$\displaystyle{\mathstrut #2}$}%
                        \setbox\tempb=\vbox{\halign{##\cr
        \mud{#1}\cr
        \noalign{\vskip\the\lineskip}
        \noalign{\hrule height 0pt}
        \phantom{$#3$}
                \rig{\vbox to 0pt{\vss\hbox to 0pt{${\; #3}$\hss}\vss}}\cr
        \noalign{\hrule}
        \noalign{\vskip\the\lineskip}
        \mud{\copy\tempa}\cr}}
                      \tempc=\wd\tempb
                      \advance\tempc by \wd\tempa
                      \divide\tempc by 2 }
\def\birule#1#2#3{{\birulehelp{#1}{#2}{#3}
                     \hbox to \wd\tempa{\hss \box\tempb \hss}}\phantom{#3}}

\def\tildeimp{\tilde{\Rightarrow}}

\def\tildeand{\tilde{\wedge}}
\def\tildeor{\tilde{\vee}}
\def\tildetop{\tilde{\top}}
\def\tildebot{\tilde{\bot}}
\def\tildefa{\tilde{\forall}}
\def\tildeex{\tilde{\exists}}
\def\ra{\rightarrow}
\def\fa{\forall}
\def\ex{\exists}
\def\lra{\longrightarrow}
\def\calA{{\cal A}}
\def\calB{{\cal B}}
\def\calC{{\cal C}}
\def\calE{{\cal E}}
\def\calF{{\cal F}}
\def\calL{{\cal L}}
\def\calM{{\cal M}}

\def\calT{{\cal T}}
\def\pred{\mbox{\it Pred\/}}
\def\nulll{\mbox{\it Null\/}}

\newcommand\VL[1]{#1}
\newcommand\VC[1]{}

\begin{document}
\title{Truth values algebras and proof normalization}
\author{Gilles Dowek}
\institute{
\'Ecole polytechnique and INRIA,\\
LIX, \'Ecole polytechnique, 91128 Palaiseau Cedex, France\\
{\tt http://lix.polytechnique.fr/$\tilde{~}$dowek/}\\
{\tt Gilles.Dowek@polytechnique.edu}}
\date{}
\maketitle
\thispagestyle{empty}

\begin{abstract}
We extend the notion of Heyting algebra to a notion of {\em truth
values algebra} and prove that a theory is consistent if and only if
it has a $\calB$-valued model for some non trivial truth values
algebra $\calB$. A theory that has a $\calB$-valued model for all
truth values algebras $\calB$ is said to be {\em super-consistent}.
We prove that super-consistency is a model-theoretic sufficient
condition for strong normalization.
\end{abstract}

\section{Introduction}

Proving that a theory has the cut elimination property has some
similarities with proving that it has a model. These similarities
appear, for instance, in the model theoretic proofs of cut
elimination, where cut elimination is obtained as a corollary of a
strengthening of the completeness theorem, expressing that if a
formula is valid in all models of a theory, then it has a {\em cut
free} proof in this theory. Such a method has been used, for instance,
by Sch\"utte, Kanger, Beth, Hintikka and Smullyan.  It has then been
used by Tait \cite{Tait66}, Prawitz \cite{Prawitz}, Takahashi
\cite{Takahashi} and Andrews \cite{Andrews} to prove cut elimination
for simple type theory. It has been generalized, more recently, by De
Marco and Lipton \cite{DeMarcoLipton} to prove cut elimination for an
intuitionistic variant of simple type theory, by Hermant
\cite{Hermant2003,Hermant2005} to prove cut elimination for classical
and intuitionistic theories in deduction modulo and by Okada
\cite{Okada} to prove cut elimination for intuitionistic linear logic.

An alternative method to prove cut elimination is to prove that all
proofs strongly normalize. Following Tait \cite{Tait67} and Girard
\cite{Girard}, this is proved by assigning a set of proofs, called a
{\em reducibility candidate}, to each formula. Here also, the proofs
have some similarities with the construction of models, except that,
in these models, the truth values $0$ and $1$ are replaced by
reducibility candidates. This analogy has been exploited in a joint
work with Werner \cite{DowekWerner}, where we have defined a notion of
reducibility candidate valued models, called {\em pre-models}, and
proved that if a theory in deduction modulo has such a model, then
it has the strong normalization property.

The fact that both cut elimination proofs and strong normalization
proofs proceed by building models raises the problem of the difference
between cut elimination and strong normalization. It is well-known
that strong normalization implies cut elimination, but what about the
converse~?  This problem can be precisely stated in deduction modulo,
where instead of using an {\em ad hoc} notion of cut for each theory
of interest, we can formulate a general notion of cut for a large
class of theories, that subsumes the usual 
{\em ad hoc} notions. This
problem has been solved by Hermant \cite{HermantThese} and
surprisingly the answer is negative: there are theories that have the
cut elimination property, but not the strong normalization property
and even not the weak normalization property.
Thus, although the model theoretic cut elimination proofs and the
strong normalization proofs both proceed by building models, 
these methods apply to 
different theories.

In this paper, we focus on the model theoretic characterization of theories
in deduction modulo that have the strong normalization property. It
has been proved in \cite{DowekWerner} that a theory has the strong
normalization property if it has a reducibility candidate valued
model. However, the usual model constructions use very little of the
properties of reducibility candidates. In particular, these
constructions seem to work independently of the chosen variant of the
closure conditions defining reducibility candidates. 
This suggests that this notion of
reducibility candidate valued model can be further generalized, by
considering an abstract notion of reducibility candidate.

Abstracting this way on the notion of reducibility candidate leads to
introduce a class of algebras, called {\em truth values algebras},
that also generalize Heyting algebras.  However there is an important
difference between truth values algebras and Heyting algebras: in a
Heyting algebra valued model the formula $P \Leftrightarrow Q$ is
valid if and only if the formulae $P$ and $Q$ have the same
denotation. In particular, all theorems have the same denotation.
This
is not necessarily the case in truth values algebra valued models
where two theorems may have different denotation.  
Thus, truth values algebra valued models are more ``intentional'' than 
Heyting algebra valued models.  
In particular, it is
possible to distinguish in the model between the computational
equivalence of formulae (the congruence of deduction modulo, or the
definitional equality of Martin-L\"of's type theory) and the provable
equivalence: the denotations of two computationally equivalent
formulae are the same, but not necessarily those of two logically
equivalent formulae.  Thus, independently of normalization, this
generalization of Heyting algebras seems to be of interest for the
model theory of deduction modulo and type theory.

We shall first introduce the notion of truth values algebra and
compare it with the notion of Heyting algebra. Then, we shall consider
plain predicate logic, define a notion of model based on
these truth values algebras and prove a soundness and a completeness
theorem for this notion of model. We shall then show that this notion
of model extends to deduction modulo. Finally, we shall strengthen the
notion of consistency into a notion of {\em super-consistency} and prove
that all super-consistent theories have the strong normalization property.
\VC{We refer to the long version of the paper for the proofs omitted
  in this abstract.} 

\section{Truth values algebras}

\subsection{Definition}

\begin{definition}[Truth values algebra]
Let $\calB$ be a set, whose elements are called 
{\em truth values}, $\calB^{+}$ be a subset of $\calB$, 
whose elements are called {\em positive} truth values,
$\calA$ and $\calE$ be subsets of $\wp(\calB)$,
$\tildetop$ and $\tildebot$ be elements of $\calB$, $\tildeimp$,
$\tildeand$, and $\tildeor$ be functions from $\calB \times \calB$ to
$\calB$, $\tildefa$ be a function from $\calA$ to $\calB$ and
$\tildeex$ be a function from $\calE$ to $\calB$.  
The structure 
$\calB = \langle \calB, \calB^{+}, \calA, \calE, \tildetop, \tildebot,
\tildeimp, \tildeand, \tildeor, \tildefa, \tildeex\rangle$ is said to
be {\em a truth value algebra} if the set 
$\calB^{+}$ is {\em closed by the intuitionistic deduction rules} 
{\em i.e.} if for all $a$, $b$, $c$ in $\calB$, $A$ in $\calA$
and $E$ in $\calE$,
\begin{enumerate}
\item if $a~\tildeimp~b \in \calB^{+}$ and $a \in \calB^{+}$ then 
$b \in \calB^{+}$, 
\item $a~\tildeimp~b~\tildeimp~a \in \calB^{+}$, 
\item
$(a~\tildeimp~b~\tildeimp~c)~\tildeimp~(a~\tildeimp~b)~\tildeimp~a~\tildeimp~c
\in \calB^{+}$, 
\item
$\tildetop \in \calB^{+}$, 
\item $\tildebot~\tildeimp~a \in \calB^{+}$, 
\item $a~\tildeimp~b~\tildeimp~(a~\tildeand~b) \in \calB^{+}$, 
\item $(a~\tildeand~b)~\tildeimp~a \in \calB^{+}$,
\item $(a~\tildeand~b)~\tildeimp~b \in \calB^{+}$,
\item $a~\tildeimp~(a~\tildeor~b) \in \calB^{+}$,
\item $b~\tildeimp~(a~\tildeor~b) \in \calB^{+}$,
\item
$(a~\tildeor~b)~\tildeimp~(a~\tildeimp~c)~\tildeimp~(b~\tildeimp~c)~\tildeimp~  c \in \calB^{+}$, 
\item the set $a~\tildeimp~A = \{a~\tildeimp~e~|~e \in A\}$ is in
$\calA$ and the set $E~\tildeimp~a = \{e~\tildeimp~a~|~e \in E\}$ is in
$\calA$, 
\item if all elements of $A$ are in $\calB^{+}$ 
then $\tildefa~A \in \calB^{+}$, 
\item 
$\tildefa~(a~\tildeimp~A)~\tildeimp~a~\tildeimp~(\tildefa~A) \in 
\calB^{+}$, 
\item if $a \in A$, then $(\tildefa~A)~\tildeimp~a \in \calB^{+}$, 
\item if $a \in E$, then $a~\tildeimp~(\tildeex~E) \in \calB^{+}$, 
\item 
$(\tildeex~E)~\tildeimp~\tildefa~(E~\tildeimp~a)~\tildeimp~a \in \cal
B^{+}$.
\end{enumerate}
\end{definition}

\begin{definition}[Full]
A truth values algebra is said to be {\em full} if $\calA = \calE =
\wp(\calB)$, {\em i.e.} if $\tildefa~A$ and $\tildeex~A$ exist for all
subsets $A$ of $\calB$.
\end{definition}

\begin{definition}[Trivial]
A truth values algebra is said to be {\em trivial} if 
$\calB^{+} = \calB$.
\end{definition}

\begin{example}
Let $\calB = \{0,1\}$. Let $\calB^{+} = \{1\}$, $\calA = \calE =
\wp(\calB)$, $\tildetop = 1$, $\tildebot = 0$, $\tildeimp$,
$\tildeand$, $\tildeor$ be the usual boolean operations, $\tildefa$ be
the function mapping the sets $\{0\}$ and $\{0,1\}$ to $0$ and
$\varnothing$ and $\{1\}$ to $1$ and $\tildeex$ be the function
mapping the sets $\varnothing$ and $\{0\}$ to $0$ and $\{1\}$ and
$\{0,1\}$ to $1$.  Then $\langle \calB, \calB^{+}, \calA, \calE,
\tildetop, \tildebot, \tildeimp, \tildeand, \tildeor, \tildefa,
\tildeex\rangle$ is a truth value algebra.
\end{example}

\begin{example}
Let $\calB$ be an arbitrary set, $\calB^{+} = \calB$, $\calA = \calE =
\wp(\calB)$ and 
$\tildetop$, $\tildebot$, $\tildeimp$,
$\tildeand$, $\tildeor$, $\tildefa$ and
$\tildeex$ be arbitrary operations.
Then $\langle \calB, \calB^{+}, \calA, \calE,
\tildetop, \tildebot, \tildeimp, \tildeand, \tildeor, \tildefa,
\tildeex\rangle$ is a trivial truth value algebra.

\end{example}

\subsection{Pseudo-Heyting algebras}

In this section, we show that truth values algebras can
alternatively be characterized as pseudo-Heyting algebras. 

\begin{definition}[Pseudo-Heyting algebra]

Let $\calB$ be a set, $\leq$ be a relation on $\calB$, 
$\calA$ and $\calE$ be subsets of $\wp(\calB)$,
$\tildetop$ and $\tildebot$ be elements of $\calB$, $\tildeimp$,
$\tildeand$, and $\tildeor$ be functions from $\calB \times \calB$ to
$\calB$, $\tildefa$ be a function from $\calA$ to $\calB$ and
$\tildeex$ be a function from $\calE$ to $\calB$, 
the structure 
$\calB = \langle \calB, {\leq}, \calA, \calE, \tildetop, \tildebot,
\tildeimp, \tildeand, \tildeor, \tildefa, \tildeex\rangle$ is said to
be a {\em pseudo-Heyting algebra} if 
for all $a$, $b$, $c$ in $\calB$, $A$ in $\calA$ and $E$ in $\calE$,
(the relation $\leq$ is a pre-order)
\begin{itemize}
\item $a \leq a$,
\item if $a \leq b$  and $b \leq c$ then $a \leq c$,
\end{itemize}
($\tildetop$ and $\tildebot$ are maximum and minimum elements
(notice that these need not be unique))
\begin{itemize}
\item $a \leq \tildetop$,
\item $\tildebot \leq a$,
\end{itemize}
($a~\tildeand~b$ is a greatest lower bound of $a$ and $b$ and 
and $a~\tildeor~b$ is a least upper bound of $a$ and $b$ (again,
these need not be unique))
\begin{itemize}
\item $a~\tildeand~b \leq a$,
\item $a~\tildeand~b \leq b$,
\item if $c \leq a$ and $c \leq b$ then $c \leq a~\tildeand~b$,
\item $a \leq a~\tildeor~b$,
\item $b \leq a~\tildeor~b$,
\item if $a \leq c$ and $b \leq c$ then $a~\tildeor~b \leq c$,
\end{itemize}
(the set $\calA$ and $\calE$ have closure conditions)
\begin{itemize}
\item $a~\tildeimp~A$ and $E~\tildeimp~a$ are in $\calA$,
\end{itemize}
($\tildefa$ and $\tildeex$ are infinite greatest lower bound and least
upper bound)
\begin{itemize}
\item if $a \in A$ then $\tildefa~A \leq a$, 
\item if for all $a$ in $A$, $b \leq a$ then $b \leq~\tildefa~A$, 
\item if $a \in E$ then $a \leq~\tildeex~E$, 
\item if for all $a$ in $E$, $a \leq b$ then $\tildeex~E \leq b$,
\end{itemize}
and
\begin{itemize}
\item $a \leq b~\tildeimp~c$ if and only if $a~\tildeand~b \leq c$.
\end{itemize}
\end{definition}

\VL{\begin{proposition}
Consider a truth values algebra 
$\langle \calB, \calB^{+}, \calA, \calE, \tildetop, \tildebot, \tildeimp,
\tildeand, \tildeor, \tildefa, \tildeex\rangle$ 
then the algebra 
$\langle \calB, \leq, \calA, \calE, \tildetop, \tildebot, \tildeimp,
\tildeand, \tildeor, \tildefa, \tildeex \rangle$ 
where the relation $\leq$ is defined by
$a \leq b$ if and only if $a~\tildeimp~b \in \calB^{+}$ is a pseudo-Heyting
algebra.
\end{proposition}

\proof{
\begin{itemize}
\item Using 1. 2. and 3. we get that $a~\tildeimp~a \in \calB^{+}$. 

\item Using 1. 2. and 3. we get that 
$a~\tildeimp~b \in \calB^{+}$ and $b~\tildeimp~c \in \calB^{+}$ 
then $a~\tildeimp~c \in \calB^{+}$. 

\item Using 1. 2. and 4. we get that $a~\tildeimp~\tildetop \in \calB^{+}$.

\item $\tildebot~\tildeimp~a \in \calB^{+}$ is condition 5.

\item $(a~\tildeand~b)~\tildeimp~a \in \calB^{+}$ is condition 7.

\item $(a~\tildeand~b)~\tildeimp~b \in \calB^{+}$ is condition 8.

\item Using 1. 2. 3. and 6. we get that 
if $c~\tildeimp~a \in \calB^{+}$ and $c~\tildeimp~b \in \calB^{+}$ 
then $c~\tildeimp~(a~\tildeand~b) \in \calB^{+}$, 

\item $a~\tildeimp~(a~\tildeor~b) \in \calB^{+}$ is condition 9.

\item $b~\tildeimp~(a~\tildeor~b) \in \calB^{+}$ is condition 10.

\item Using 1. 2. 3. and 11. we get that if $a~\tildeimp~c \in \cal
B^{+}$ and $b~\tildeimp~c \in \calB^{+}$ then
$(a~\tildeor~b)~\tildeimp~c \in \calB^{+}$.

\item The closure conditions are 12.

\item If $a~\in A$ then $(\tildefa~A)~\tildeimp~a \in \calB^{+}$ is
  condition 15.

\item From 13. and 14. we get if for all $a~\in A$ $b~\tildeimp~a  \in
\calB^{+}$, then $b~\tildeimp~\tildefa~A \in \calB^{+}$, 

\item If $a~\in A$ then $a~\tildeimp~(\tildeex~A) \in \calB^{+}$ is
  condition 16.

\item From 1. 2. 3. 13. and 17. we get that if for all $a~\in A$
$a~\tildeimp~b  \in \calB^{+}$, then $\tildeex~A~\tildeimp~b \in
\calB^{+}$,  

\item From 1. 2. 3. 6. 7. and 8. we get that
  $a~\tildeimp~(b~\tildeimp~c) \in \calB^{+}$ if and only if
  $(a~\tildeand~b)~\tildeimp~c \in \calB^{+}$. 
\end{itemize}}}

\VL{\begin{proposition}
Consider a pseudo-Heyting algebra $\langle \calB, \leq, \calA, \calE,
\tildetop, \tildebot, \tildeimp, \tildeand, \tildeor, \tildefa,
\tildeex \rangle$, then the algebra $\langle \calB, \calB^{+}, \calA,
\calE, \tildetop, \tildebot, \tildeimp, \tildeand, \tildeor, \tildefa,
\tildeex \rangle$, where $\calB^{+} = \{x~|~\tildetop \leq x\}$ is a truth
values algebra.
\end{proposition}

\proof{Let us first prove the following lemma: if $x \leq
a~\tildeimp~b$ and $x \leq a$ then $x \leq b$. From $x \leq x$ and $x
\leq a$, we get $x \leq x~\tildeand~a$ and as $x \leq a~\tildeimp~b$,
we have $x~\tildeand~a \leq b$. By transitivity, we get $x \leq b$. 

\begin{enumerate}
\item 
Using the lemma above, we get 
that if $\tildetop \leq a~\tildeimp~b$ and $\tildetop \leq a$
then $\tildetop \leq b$.

\item We have 
$(\tildetop~\tildeand~a)~\tildeand~b \leq 
\tildetop~\tildeand~a \leq a$.
Hence $\tildetop \leq a~\tildeimp~b~\tildeimp~a$.

\item Let $x = \tildetop~\land~(a~\tildeimp~b~\tildeimp~c)~\tildeand~
(a~\tildeimp~b)~\tildeand~a$. 
We have $x \leq a~\tildeimp~b~\tildeimp~c$,
$x \leq a~\tildeimp~b$
and $x \leq a$. 
Using the lemma three times, we get 
$x \leq c$. Hence $\tildetop \leq
(a~\tildeimp~b~\tildeimp~c)~\tildeimp~(a~\tildeimp~b)~\tildeimp~a~\tildeimp~c$.

\item We have $\tildetop \leq \tildetop$. 

\item We have $\tildetop~\tildeand~\tildebot \leq \tildebot \leq a$, thus 
$\tildetop~\tildeand~\tildebot \leq a$. Hence 
$\tildetop \leq \tildebot~\tildeimp~a$. 

\item We have $(\tildetop~\tildeand~a)~\tildeand~b \leq
  \tildetop~\tildeand~a \leq a$ and
$(\tildetop~\tildeand~a)~\tildeand~b \leq b$.
Hence 
$(\tildetop~\tildeand~a)~\tildeand~b \leq a~\tildeand~b$
and
$\tildetop \leq a~\tildeimp~b~\tildeimp~(a~\tildeand~b)$.

\item We have $\tildetop~\tildeand~(a~\tildeand~b) \leq
(a~\tildeand~b) \leq a$. 
Hence $\tildetop \leq (a~\tildeand~b)~\tildeimp~a$.

\item We have $\tildetop~\tildeand~(a~\tildeand~b) \leq
(a~\tildeand~b) \leq b$. 
Hence $\tildetop \leq (a~\tildeand~b)~\tildeimp~b$.

\item We have $\tildetop~\tildeand~a \leq a \leq 
(a~\tildeor~b)$. 
Hence $\tildetop \leq a~\tildeimp~(a~\tildeor~b)$.

\item We have $\tildetop~\tildeand~b \leq b \leq 
(a~\tildeor~b)$. 
Hence $\tildetop \leq b~\tildeimp~(a~\tildeor~b)$.

\item
Let $x =
\tildetop~\tildeand~(a~\tildeor~b)~\tildeand~(a~\tildeimp~c)~\tildeand~(b~\tildeimp~c)$.
We have $x \leq a~\tildeor~b$, 
$x \leq a~\tildeimp~c$ and
$x \leq b~\tildeimp~c$.
We have $x \leq a~\tildeor~b$ and $x \leq x$, hence 
$x \leq (a~\tildeor~b)~\tildeand~x$. 
We have $a~\tildeand~x \leq x~\tildeand~a \leq c$, thus 
$a~\leq x~\tildeimp~c$ and, in a similar way,
$b~\leq x~\tildeimp~c$. Thus, we have 
$a~\tildeor~b \leq x~\tildeimp~c$,
{\em i.e.} 
$(a~\tildeor~b)~\tildeand~x \leq c$. By transitivity, we conclude $x
\leq c$. Hence $\tildetop \leq (a~\tildeor~b)~\tildeimp~(a~\tildeimp~c)~\tildeimp~(b~\tildeimp~c)~\tildeimp~c$.

\item The closure conditions of $\calA$ and $\calE$ are the same.

\item
If all elements of $A$ are in $\calB^{+}$ then 
$\tildefa~A \in \calB^{+}$. Indeed, for 
all elements $x$ of $A$, $\tildetop \leq x$ hence
$\tildetop \leq \tildefa~A$. 

\item
Let 
$x = \tildetop~\land~\tildefa~(a~\tildeimp~A)~\tildeand~a$.
Let $y$ be an arbitrary element of $A$. Notice that, by definition
of $a~\tildeimp~A$, 
$a~\tildeimp~y \in a~\tildeimp~A$. 
We have $x \leq \tildefa~(a~\tildeimp~A)$ hence 
$x \leq a~\tildeimp~y$. We also have $x \leq a$, hence using the lemma
$x \leq y$. 

For all $y \in A$, we have $x \leq y$, thus $x \leq \fa~A$. 
Hence $\tildetop \leq
\tildefa~(a~\tildeimp~A)~\tildeimp~a~\tildeimp~(\tildefa~A)$. 

\item If $a \in A$, then we have $\tildetop~\tildeand~(\tildefa~A)
\leq \tildefa~A \leq a$. Hence $\tildetop \leq (\tildefa~A)~\tildeimp~a$.

\item If $a \in A$, then we have $\tildetop~\tildeand~a
\leq a \leq \tildeex~A$. Hence $\tildetop \leq
a~\tildeimp~(\tildeex~A)$.

\item Let $x =
\tildetop~\tildeand~(\tildeex~A)~\tildeand~\tildefa~(A~\tildeimp~a)$. 
We have 
$x \leq (\tildeex~A)$ and 
$x \leq \tildefa~(A~\tildeimp~a)$. 
We have $x \leq (\tildeex~A)$ and $x \leq x$ thus
$x \leq (\tildeex~A)~\tildeand~x$. 
As $x \leq \tildefa~(A~\tildeimp~a)$. For all $y$ in $A$, $x \leq
y~\tildeimp~a$, {\em i.e} $(x~\tildeand~y) \leq a$. 
For all $y$ in $A$, we have $y~\tildeand~x \leq x~\tildeand~y \leq
a$ {\em i.e.} $y \leq x~\tildeimp~a$. 
Thus, $\tildeex~A \leq x~\tildeimp~a$, {\em i.e.} $\tildeex~A~\tildeand~x
\leq a$. By transitivity, we get $x \leq a$. Hence 
$\tildetop \leq
(\tildeex~A)~\tildeimp~\tildefa~(A~\tildeimp~a)~\tildeimp~a$.
\end{enumerate}}}

\VC{\begin{proposition}
Consider a truth values algebra 
$\langle \calB, \calB^{+}, \calA, \calE, \tildetop, \tildebot, \tildeimp,
\tildeand, \tildeor, \tildefa, \tildeex\rangle$ 
then the algebra 
$\langle \calB, \leq, \calA, \calE, \tildetop, \tildebot, \tildeimp,
\tildeand, \tildeor, \tildefa, \tildeex \rangle$ 
where the relation $\leq$ is defined by
$a \leq b$ if and only if $a~\tildeimp~b \in \calB^{+}$ is a pseudo-Heyting
algebra.

Conversely, consider a pseudo-Heyting algebra $\langle \calB, \leq,
\calA, \calE, \tildetop, \tildebot, \tildeimp, \tildeand, \tildeor,
\tildefa, \tildeex \rangle$, then the algebra $\langle \calB,
\calB^{+}, \calA, \calE, \tildetop, \tildebot, \tildeimp, \tildeand,
\tildeor, \tildefa, \tildeex \rangle$, where $\calB^{+} = \{x~|~\tildetop
\leq x\}$ is a truth values algebra.
\end{proposition}

}

\begin{definition}[Heyting algebra]
A pseudo-Heyting algebra is said to be a {\em Heyting algebra} if
the relation $\leq$ is antisymmetric
\begin{itemize}
\item $x \leq y  \Rightarrow y \leq x \Rightarrow x = y$.
\end{itemize}
\end{definition}

\noindent {\em Remark.}
If the pseudo-Heyting algebra 
$\langle \calB, \leq, \calA, \calE, \tildetop, \tildebot, \tildeimp,
\tildeand, \tildeor, \tildefa, \tildeex\rangle$ is a Heyting algebra,
then the set $\calB^{+} = \{x~|~\tildetop \leq x\}$ is the singleton 
$\{\tildetop\}$.
Indeed, if $a \in \calB^{+}$ then $\tildetop \leq a$ and $a \leq
\tildetop$. Hence $a = \tildetop$.

\begin{definition}
A function $F$ from a truth value algebra $\calB_1$ to 
a truth value algebra $\calB_2$ is said to be {\em a morphism of 
truth values algebras} if 
\begin{itemize}
\item $x \in \calB^{+}_1$ if and only if $F(x) \in \calB^{+}_2$, 
\item if $A \in \calA_1$ then $F(A) \in \calA_2$,
if $E \in \calE_1$ then $F(E) \in \calE_2$,
\item $F(\tildetop_1) = \tildetop_2$, $F(\tildebot_1) = \tildebot_2$, 
$F(a~\tildeimp_1~b) = F(a)~\tildeimp_2~F(b)$,
$F(a~\tildeand_1~b) = F(a)~\tildeand_2~F(b)$,
$F(a~\tildeor_1~b) = F(a)~\tildeor_2~F(b)$,
$F(\tildefa_1~A) = \tildefa_2~F(A)$, 
$F(\tildeex_1~E) = \tildeex_2~F(E)$.
\end{itemize}
{\em Morphisms of pseudo-Heyting algebras} are defined in a similar way
except that the first condition is replaced by 
\begin{itemize}
\item $x \leq_1 y$ if and only if $F(x) \leq_2 F(y)$.
\end{itemize}
\end{definition}

\begin{proposition}
\label{heyting}
Let $\calB$ be a pseudo-Heyting algebra, then there exists a 
pseudo-Heyting algebra $\calB/\calB^{+}$ that is a Heyting algebra and a 
morphism of pseudo-Heyting algebras $\Phi$ from $\calB$ to
$\calB/\calB^{+}$.
\end{proposition}

\VL{\proof{We define a relation $\simeq$ on elements of $\calB$ by 
$a \simeq b$ if and only if $a \leq b$ and $b \leq a$. It is routine to check
that this relation is an equivalence relation and that all the
operations of $\calB$ are compatible with this relation. We define
$\calB/\calB^{+}$ as the quotient $\calB / \simeq$ and the morphism $\Phi$
by $\Phi(a) = a / \simeq$.}
\medskip}

\noindent {\em Remark.}
We have proved that, in the definition of Heyting algebras, the
antisymmetry is useless and can be dropped.  The equivalence of truth
values algebras and pseudo-Heyting algebras shows that antisymmetry is
the only property that can be dropped and that truth values algebras
are, in some sense, the best possible generalization of Heyting
algebras, as we cannot require less than closure by intuitionistic
deduction rules.

\subsection{Examples of truth values algebras}

We have seen that the algebra $\{0,1\}$ is a truth value algebra and
more generally that all Heyting algebras are truth values algebras.
We give in this section two examples of truth values algebras that are
not Heyting algebras.

\begin{example}
The truth value algebra $\calT_1$ is defined as follows. 
The set $\calT_1$ is $\{0, I, 1\}$ and the set $\calT_1^{+}$
is $\{I, 1\}$. The sets $\calA$ and $\calE$ are $\wp(\calT_1)$.
The functions $\tildetop$, $\tildebot$,
$\tildeand$, $\tildeor$, $\tildefa$ and $\tildeex$ are the same as in
the algebra $\{0,1\}$, except that their value on $I$ is the same as
their value on $1$. For instance the table of the operation $\tildeor$ is 
$$\begin{array}{|c|ccc|}
\hline
& 0 & I & 1\\
\hline
0     &  0 & 1 & 1\\
I     &  1 & 1 & 1\\
1     &  1 & 1 & 1\\
\hline
\end{array}$$
The function $\tilde{\Rightarrow}$ is defined by the table 
$$\begin{array}{|c|ccc|}
\hline
      & 0 & I & 1\\
\hline
0     &  1 & 1 & 1\\
I     &  0 & 1 & 1\\
1     &  0 & I & I\\
\hline
\end{array}$$
Notice that as $I~\tildeimp~1$ and $1~\tildeimp~I$ are both in
$\calT_1^{+}$ we have $I \leq 1$ and $1 \leq I$. Hence the relation
$\leq$ is not antisymmetric and the truth value algebra $\calT_{1}$ is
not a Heyting algebra.
\end{example}

\begin{example} 
The truth value algebra $\calT_2$ is similar to $\calT_1$,
except that the function 
$\tilde{\Rightarrow}$ is defined by the table 
$$\begin{array}{|c|ccc|}
\hline
      & 0 & I & 1\\
\hline
0     &  1 & 1 & I\\
I     &  0 & 1 & I\\
1     &  0 & 1 & I\\
\hline
\end{array}$$
\end{example}

\subsection{Ordered truth values algebras}

We consider truth values algebras extended with an order
relation $\sqsubseteq$ on $\calB$. This order relation extends to sets of truth
values in a trivial way: $A \sqsubseteq B$ if for all $x$ in $A$ there
exists a $y$ in $B$ such that $x \sqsubseteq y$.

\begin{definition}[Ordered truth values algebra]
An {\em ordered truth values algebra} is a 
truth values algebra together with a relation 
$\sqsubseteq$ on $\calB$ such that 
\begin{itemize}
\item $\sqsubseteq$ is an order relation,
\item $\calB^{+}$ is upward closed,
\item $\tildetop$ is a maximal element, 
\item $\tildeand$, $\tildeor$, 
$\tildefa$ and $\tildeex$ are monotonous, 
$\tildeimp$ is left anti-monotonous and right monotonous.
\end{itemize}
\end{definition}

\begin{definition}[Complete ordered truth values algebra]
A ordered truth values algebra is said to be {\em complete} if
every subset of $\calB$ has a greatest lower bound for $\sqsubseteq$.
Notice that this implies that every subset also has a least upper
bound. We write $glb(a,b)$ and $lub(a,b)$ the greatest lower bound and
the least upper bound of $a$ and $b$ for the order $\sqsubseteq$.
\end{definition}

\begin{example}
The algebra $\calT_1$ ordered by $0 \sqsubseteq I
\sqsubseteq 1$ is complete. 
\end{example}

\begin{example}
The algebra $\calT_2$ cannot be extended to a complete ordered
algebra. Indeed the set $\{I, 1\}$ would need to have a least upper
bound. This least upper bound cannot be $0$ because $\calT_2^{+}$
would then not be upward closed. If it were $1$ then we would have $I
\sqsubseteq 1$ and thus $1~\tildeimp~I \sqsubseteq 1~\tildeimp~1$,
{\em i.e.} $1 \sqsubseteq I$. Thus the relation $\sqsubseteq$ would
not be antisymmetric. 
If it were $I$ then we would have $1
\sqsubseteq I$ and thus $1~\tildeimp~1 \sqsubseteq 1~\tildeimp~I$,
{\em i.e.} $I \sqsubseteq 1$. Thus the relation $\sqsubseteq$ would
not be antisymmetric. 
\end{example}

\VL{
\begin{proposition}\label{finer}
The order relation $\sqsubseteq$ is finer than
$\leq$, {\em i.e.} if $a \sqsubseteq b$ then $a \leq b$. 
\end{proposition}

\proof{If $a \sqsubseteq b$, then $a~\tildeimp~a \sqsubseteq
a~\tildeimp~b$, hence $a~\tildeimp~b \in \calB^{+}$, {\em i.e.} $a
\leq b$.}

\begin{proposition}\label{almost}
$$glb(a,b) \leq a~\tildeand~b \leq glb(a~\tildeand~\tildetop,
\tildetop~\tildeand~b)$$
\end{proposition}

\proof{We have $glb(a,b) \sqsubseteq a$ and $glb(a,b) \sqsubseteq
b$. Thus, by Proposition \ref{finer}, $glb(a,b) \leq a$ and $glb(a,b)
\leq b$. Thus $glb(a,b) \leq a~\tildeand~b$.

We have $a \sqsubseteq a$ and $b \sqsubseteq \tildetop$ thus
$a~\tildeand~b \sqsubseteq a~\tildeand~\tildetop$.  Similarly,
$a~\tildeand~b \sqsubseteq \tildetop~\tildeand~b$.  Thus
$a~\tildeand~b \sqsubseteq
glb(a~\tildeand~\tildetop,\tildetop~\tildeand~b)$ and, by Proposition
\ref{finer}, $a~\tildeand~b \leq
glb(a~\tildeand~\tildetop,\tildetop~\tildeand~b)$.}
}
\begin{proposition}
In a Heyting algebra, $\leq$ and $\sqsubseteq$ are extensionally
equal, {\em i.e.} $a \sqsubseteq b$ if and only if $a \leq b$. 
\end{proposition}

\VL{\proof{In a Heyting algebra the relation $\leq$ is antisymmetric and 
$a = a~\tildeand~\tildetop = \tildetop~\tildeand~a$. Thus, from
Proposition \ref{almost}, we get $glb(a,b) = a~\tildeand~b$. If $a
\leq b$, we have $a~\tildeand~b = a$, thus 
$glb(a,b) = a$, thus $a \sqsubseteq b$. Conversely, by Proposition
\ref{finer}, if $a \sqsubseteq b$ then $a \leq b$.}}

\subsection{Completion}

We now want to prove that for any truth value algebra $\calB$,
there is another truth value algebra $\calB_C$ that is full, ordered
and complete and a morphism $\Phi$ from $\calB$ to $\calB_C$. 
Notice that we do not require the morphism $\Phi$ to be injective. 

There are two ways to prove this, the first is to use Proposition
\ref{heyting} in a first step to build a truth value algebra $\calB/\calB^{+}$
that is a Heyting algebra and a morphism for $\calB$ to $\calB/\calB^{+}$ and
then apply in a second step MacNeille completion to the algebra
$\calB/\calB^{+}$ to embed it into a full Heyting algebra. Together with its
natural order, this algebra is a full, ordered and complete truth
value algebra.
The second is to apply MacNeille completion directly to $\calB$
noticing that antisymmetry is not used in MacNeille completion, except
to prove the injectivity of the morphism. \VC{The proof is detailed in
  the long version of the paper.}

\VL{
To keep the paper
self-contained we follow this second way.

\begin{definition}[Closure]
Let $X$ a subset of $\calB$, then the set of upper bounds of $X$ is 
$u(X) = \{y~|~\fa x~(x \in X \Rightarrow x \leq y)\}$
the set of lower bounds of $X$ is 
$l(X) = \{y~|~\fa x~(x \in X \Rightarrow y \leq x)\}$
and the closure of $X$ is $C(X) = l(u(X)$.
\end{definition}

\VL{It is easily checked that
$$X \subseteq Y \Rightarrow u(Y) \subseteq u(X)$$
$$X \subseteq Y \Rightarrow l(Y) \subseteq l(X)$$
$$X \subseteq Y \Rightarrow C(X) \subseteq C(Y)$$

\begin{proposition}
$X \subseteq C(X)$
\end{proposition}

\proof{Consider $x \in X$. For all $y \in u(X)$, $x \leq y$. Hence 
$x \in l(u(X))$.}

\begin{proposition}
\label{exist}
If $X \in \calE$, then $\tildeex~X \in C(X)$
\end{proposition}

\proof{For all $y \in u(X)$, $\tildeex~X \leq y$. Hence 
$\tildeex~X \in l(u(X))$.}}

\begin{definition}[Closed]
A subset $X$ of $\calB$ is said to be closed if $C(X) = X$\VL{or,
equivalently, $C(X) \subseteq X$}. 
\end{definition}

\VL{\begin{proposition}
Any set of the form $C(X)$ is closed. 
\end{proposition}

\proof{Consider an arbitrary set $Z$. Consider $z \in Z$. 
For all $y \in l(Z)$, $y \leq z$. Hence 
$z \in u(l(Z))$. Thus, for an arbitrary $Z$, $Z \subseteq u(l(Z))$ and
in particular $u(X) \subseteq u(l(u(X)))$. Thus $l(u(l(u(X))))
\subseteq l(u(X))$, {\em i.e.} $C(C(X)) \subseteq C(X)$.}

\begin{proposition}
$$a \in C(\{b\}) \Leftrightarrow a \leq b$$
$$C(\{a\}) \subseteq C(\{b\}) \Leftrightarrow a \leq b$$
\begin{center}
If $X$ is closed, $x \in X$ and $y \leq x$ then $y \in X$
\end{center}
\end{proposition}

\proof{
$a \in C(\{b\})$ if and only if $a \in l(u(\{b\}))$ if and only if
$\fa x~(x \in u(\{b\}) \Rightarrow a \leq x)$ if and only if $\fa x~(b \leq x
\Rightarrow a \leq x)$ if and only if (as the relation $\leq$ is
reflexive and transitive) $a \leq b$.

If $C(\{a\}) \subseteq C(\{b\})$ then $\{a\} \subseteq C(\{a\})
\subseteq C(\{b\})$. Hence $a \in C(\{b\})$, {\em i.e.} $a \leq
b$. Conversely, if $a \leq b$ then $a \in C(\{b\})$, 
$\{a\} \subseteq C(\{b\})$, hence 
$C(\{a\}) \subseteq C(C(\{b\})) = C(\{b\})$.

If $X$ is closed, $x \in X$ and $y \leq x$ then let $z$ be an element
of $u(X)$, we have $x \leq z$ and $y \leq x$, thus, by transitivity,
$y \leq z$, {\em i.e.} $y \in l(u(X)) = C(X) = X$.}
}

\begin{proposition}
Let $\calB$ be a pseudo-Heyting algebra.  Let $\calB_C$ be the set of
closed subsets of $\calB$, 
$\leq_{C}$ be inclusion, 
$\calA_C = \calE_C = \wp(\calB_C)$,
$\tildebot_C = C(\{\tildebot\})$,
$\tildetop_C = C(\{\tildetop\})$, $\tildeand_C$ be intersection,
$\tildeor_C$ be defined by $X \tildeor_{C} Y = C(X \cup Y)$,
$\tildefa_C$ be intersection, $\tildeex_C$ be defined by
$\tildeex_{C} E = C(\bigcup E)$, $\tildeimp_{C}$ be defined by $X
\tildeimp_{C} Y = \bigcap_{C(\{x\}) \subseteq X, Y \subseteq C(\{y\})}
C(\{x~\tildeimp~y\})$.  

Then the structure $\langle \calB_C, \leq_C, \calA_C, \calE,
\tildebot_C, \tildetop_C, \tildeimp_C, \tildeand_C, \tildeor_C,
\tildefa_C, \tildeex_C \rangle$ is a full, ordered and complete
Heyting algebra. 
\end{proposition}

\VL{\proof{\begin{enumerate}
\item Inclusion is trivially an order relation. 

\item The set $C(\{\tildebot\})$ is closed.
It is a minimum in $\calB_C$, because if $X$ is closed, it is a set of
lower bounds and hence $\tildebot \in X$. Thus $C(\{\tildebot\})
\subseteq C(X) = X$. 

\item The set $C(\{\tildetop\})$ is closed. It is a maximum in
$\calB_C$, because it is equal to $\calB$.

\item Let us check that binary and arbitrary intersections of closed
sets are closed sets. We detail only the case of the operation 
$\tildefa$, the binary operation being a particular case.
If all elements of $A$ are closed sets, we have 
for every $X$ in $A$, 
$\tildefa_{\calC}~A \subseteq X$, $C(\tildefa_{\calC}~A) \subseteq
C(X) = X$. Thus
$C(\tildefa_{\calC}~A) \subseteq \tildefa_{\calC}~A$.  Moreover, binary and
arbitrary intersections are obviously greatest lower bounds in $\calB_C$.

\item The sets $C(X \cup Y)$ and 
$C(\bigcup E)$ are closed. Let us check that they are
least upper bounds. Again, we detail case of the case of the operation
$\tildeex_C$, the binary operation being a particular case. The
operation $\tildeex_C$ is an upper bound as if $X$ is an element of $E$, 
$X \subseteq \bigcup E \subseteq C(\bigcup E)$.  
Then, it is the least as, if $A$ is an element of $\calB_{\calC}$ 
such that for all $X$ in $E$, 
$X \subseteq Z$ then $\bigcup E \subseteq Z$ hence $C(\bigcup E)
\subseteq C(Z)$ and $C(\bigcup E) \subseteq Z$.

\item 
The set $\bigcap_{C(\{x\}) \subseteq X, Y \subseteq C(\{y\})} C(\{x~
\tildeimp~y\})$ is closed as it is an intersection of closed sets.
Let us check that $X \leq_C A~\tildeimp_C~B$ if and only if
$X~\tildeand_C~A \subseteq B$.  

Assume $X \subseteq A~\tildeimp_C~B$ and let $x \in X~\tildeand_C~A$.
Let $b \in u(B)$. 
We have $x \in X$, thus $x \in A~\tildeimp_C~B$.
We have $x \in A$, thus $C(\{x\}) \subseteq C(A) = A$.
We have $b \in u(B)$, thus $\fa y~(y \in B \Rightarrow y \leq b)$, 
$\fa y~(y \in B \Rightarrow y \in C(\{b\}))$ and $B \subseteq C(\{b\})$.
We have $x \in A~\tildeimp_C~B$, $C(\{x\}) \subseteq A$, 
and $B \subseteq C(\{b\})$ thus $x \in C(\{x~\tildeimp~b\})$, 
{\em i.e.} $x \leq x~\tildeimp~b$,
$x~\tildeand~x \leq b$, 
$x \leq b$. 
Thus, for all $b$ in $u(B)$, $x \leq b$, {\em i.e.}
$x \in l(u(B))$, $x \in C(B)$ and $x \in B$.

Conversely, assume $X~\tildeand_C~A \subseteq B$. Let $x \in X$.  Let
$a$ such that $C(\{a\}) \subseteq A$ and $b$ such that $B \subseteq
C(\{b\})$. We have $x~\tildeand~a \leq x$ and $x \in X$, hence
$x~\tildeand~a \in X$. We have $x~\tildeand~a \leq a \in C(\{a\})
\subseteq A$, hence $x~\tildeand~a \in A$. Thus $x~\tildeand~a \in
X~\tildeand_C~A \subseteq B \subseteq C(\{b\})$. Therefore
$x~\tildeand~a \in C(\{b\})$, $x~\tildeand~a \leq b$, $x \leq
a~\tildeimp~b$ and $x \in C(\{a~\tildeimp~b\})$.  Thus 
$x \in A~\tildeimp_C~B$.
\end{enumerate}}}

\begin{proposition}
The function $a \mapsto C(\{a\})$ is a morphism of pseudo-Heyting algebras.
\end{proposition}

\VL{\proof{
\begin{enumerate}
\item $a \leq b \Leftrightarrow C(\{a\}) \subseteq C(\{b\})$.

\item $C(\{\tildetop\}) = \tildetop_{\calC}$ by definition.

\item $C(\{\tildebot\}) = \tildebot_{\calC}$ by definition.

\item $C(\{a~\tildeand~b\}) = C(\{a\})~\tildeand_{\calC}~C(\{b\})$. Indeed 
$x \in C(\{a~\tildeand~b\})$ 
if and only if 
$x \leq a~\tildeand~b$ 
if and only if 
$x \leq a$ and $x \leq b$ 
if and only if 
$x \in C(\{a\})$ and $x \in C(\{b\})$ 
if and only if 
$x \in C(\{a\})~\tildeand_{C}~(\{b\})$.

\item $C(\{a~\tildeor~b\}) =
  C(\{a\})~\tildeor_{\calC}~C(\{b\})$. Indeed 
we have $a \in 
C(\{a\})~\cup~C(\{b\})$, thus if $x \in u(C(\{a\})~\cup~C(\{b\}))$, then
$a \leq x$.  In a similar way, 
if $x \in u(C(\{a\})~\cup~C(\{b\}))$, then $b \leq x$.  Thus 
if $x \in u(C(\{a\})~\cup~C(\{b\}))$, then $a~\tildeor~b \leq x$,
{\em i.e.}
if $x \in u(C(\{a\})~\cup~C(\{b\}))$, then $x \in
u(\{a~\tildeor~b\})$
{\em i.e.}
$u(C(\{a\})~\cup~C(\{b\})) \subseteq u(\{a~\tildeor~b\})$. 
Hence 
$l(u(\{a~\tildeor~b\})) \subseteq l(u(C(\{a\})~\cup~C(\{b\})))$
{\em i.e.}
$C(\{a~\tildeor~b\}) \subseteq C(C(\{a\})~\cup~C(\{b\})))$
{\em i.e.}
$C(\{a~\tildeor~b\}) \subseteq C(\{a\})~\tildeor_{\calC}~C(\{b\}))$.
Conversely, we have $a \leq a~\tildeor~b$, thus 
$C(\{a\}) \subseteq C(\{a~\tildeor~b\})$. 
In a similar way 
$C(\{b\}) \subseteq C(\{a~\tildeor~b\})$.
Thus 
$C(\{a\}) \cup C(\{b\}) \subseteq C(\{a~\tildeor~b\})$
and 
$C(C(\{a\}) \cup C(\{b\})) \subseteq C(C(\{a~\tildeor~b\}))$
{\em i.e.}
$C(\{a\})~\tildeor_{\calC}~C(\{b\}) \subseteq C(\{a~\tildeor~b\})$.

\item $C(\{\tildefa~A\}) = \tildefa_C~\{C(\{a\})~|~a \in A\}$.
Indeed, $x \in C(\{\tildefa~A\})$ 
if and only if 
$x \leq \tildefa~A$ 
if and only if 
for all $a$ in $A$, $x \leq a$ if and only if for all $a$ in $A$, 
$x \in C(\{a\})$ if and only if
$x \in \tildefa_C~\{C(\{a\})~|~a \in A\}$.

\item 
$C(\{\tildeex~E\}) = \tildeex_C~\{C(\{e\})~|~e \in E\}$.
Indeed, for all $e$ in $E$ we have $e \in C(\{e\})$ and thus 
$E \subseteq \bigcup \{C(\{e\})~|~e \in E\}$ and 
$C(E) \subseteq C(\bigcup \{C(\{e\})~|~e \in E\})$
{\em i.e.} $C(E) \subseteq \tildeex_C~\{C(\{e\})~|~e \in E\}$.
As, by Proposition \ref{exist}, $\tildeex~E \in C(E)$, we have
$C(\{\tildeex~E\}) \subseteq C(E)$, and thus
$C(\{\tildeex~E\}) \subseteq \tildeex_C~\{C(\{e\})~|~e \in E\}$. 
Conversely, for all $e$ in $E$, $e \leq \tildeex~E$, hence 
$C(\{e\}) \subseteq C(\{\tildeex~E\})$. Thus,
$\bigcup \{C(\{e\})~|~e \in E\} \subseteq C(\{\tildeex~E\})$
and
$C(\bigcup \{C(\{e\})~|~e \in E\}) \subseteq C(\{\tildeex~E\})$
{\em i.e.}
$\tildeex_C~\{C(\{e\})~|~e \in E\} \subseteq C(\{\tildeex~E\})$.

\item 
$C(\{a~\tildeimp~b\}) = C(\{a\})~\tildeimp_C~C(\{b\})$.
Indeed, 
$z \in C(\{a\})~\tildeimp_C~C(\{b\})$
if and only if
for all $x$ and $y$ such that 
$C(\{x\}) \subseteq C(\{a\})$ and $C(\{b\}) \subseteq C(\{y\})$ we
have $z \in C(\{x~\tildeimp~y\})$
if and only if 
for all $x$ and $y$ such that 
$x \leq a$ and $b \leq y$, we have $z \leq x~\tildeimp~y$
if and only if 
$z \leq a~\tildeimp~b$ 
if and only if 
$z \in C(\{a~\tildeimp~b\})$. 
\end{enumerate}}
}

\begin{proposition}
\label{macneille}
Let $\calB$ be a truth value algebra, then there exists a full,
ordered and complete truth value algebra $\calB_C$ and a morphism of 
truth values algebras from $\calB$ to $\calB_C$.
\end{proposition}

\VL{\proof{The Heyting algebra $\calB_C$ is a full, ordered and complete
truth value algebra and the function $a \mapsto C(\{a\})$ is a
morphism of truth values algebras.}}
}

\begin{example}
The algebra $\calT_2$ cannot be extended to a complete ordered
algebra, but it can be embedded with a non injective morphism in the
full ordered and complete algebra $\{0,1\}$. 
\end{example}

\section{Predicate Logic}

\subsection{Models}

\begin{definition}[$\calB$-valued structure]
Let $\calL = \langle f_i, P_j \rangle$ be a language in predicate
logic and $\calB$ be a truth values algebra, a {\em $\calB$-valued structure}
for the language $\calL$, $\calM = \langle \calM, \calB, \hat{f}_i,
\hat{P}_j \rangle$ is a structure such that $\hat{f_i}$ is a function
from $\calM^n$ to $\calM$ where $n$ is the arity of the symbol $f_i$
and $\hat{P_j}$ is a function from $\calM^n$ to $\calB$ where $n$ is
the arity of the symbol $P_i$.

This definition extends trivially to many-sorted languages.
\end{definition}

\begin{definition}[Denotation]
Let $\calB$ be a truth values algebra, $\calM$ be a $\calB$-valued structure
and $\phi$ be an assignment. The denotation $\llbracket A
\rrbracket_\phi$ of a formula $A$ in $\calM$ is defined as follows

\begin{itemize}
\item $\llbracket x \rrbracket_{\phi} = \phi(x)$, 
\item $\llbracket f(t_1, ..., t_n) \rrbracket_{\phi} = 
\hat{f}(\llbracket t_1\rrbracket_{\phi}, ..., \llbracket
t_n\rrbracket_{\phi})$,
\item $\llbracket P(t_1, ..., t_n) \rrbracket_{\phi} = 
\hat{P}(\llbracket t_1\rrbracket_{\phi}, ..., \llbracket
t_n\rrbracket_{\phi})$,
\item $\llbracket \top \rrbracket_{\phi} = \tildetop$, 
\item $\llbracket \bot \rrbracket_{\phi} = \tildebot$, 
\item $\llbracket A \Rightarrow B \rrbracket_{\phi} = 
\llbracket A \rrbracket_{\phi} 
~\tildeimp~
\llbracket B \rrbracket_{\phi}$, 
\item
$\llbracket A \wedge B \rrbracket_{\phi} = 
\llbracket A \rrbracket_{\phi}
~\tildeand~
\llbracket B \rrbracket_{\phi}$, 
\item
$\llbracket A \vee B \rrbracket_{\phi} = 
\llbracket A \rrbracket_{\phi}
~\tildeor~
\llbracket B \rrbracket_{\phi}$, 
\item $\llbracket \fa x~A \rrbracket_{\phi} = 
\tildefa~\{
\llbracket A \rrbracket_{\phi + \langle x, e\rangle}~|~e \in \calM\}$,
\item $\llbracket \ex x~A \rrbracket_{\phi} = 
\tildeex~\{
\llbracket A \rrbracket_{\phi + \langle x, e\rangle}~|~e \in \calM\}$.
\end{itemize}
Notice that the denotation of a formula containing quantifiers may be
undefined, but it is always defined if the truth value algebra is
full.
\end{definition}

\begin{definition}[Model]
A formula $A$ is said to be {\em valid} in a $\calB$-valued structure 
$\calM$, and the $\calB$-valued structure $\calM$ is said to be {\em a
model of} $A$, $\calM \models A$, if for all
assignments $\phi$,  
$\llbracket A \rrbracket_\phi$ is defined and is a positive truth
value.

The $\calB$-valued structure
$\calM$ is said to be {\em a model of} a theory 
$\calT$ if it is a model of all the axioms of $\calT$.
\end{definition}

\subsection{Soundness and completeness}

As the notion of truth values algebra extends that of Heyting algebra,
the completeness theorem for the notion of model introduced above is a
simple corollary of the completeness theorem for the notion of model
based on Heyting algebras.  But, it has a simpler direct proof. 
It is well-known that completeness proofs for boolean algebra valued
models and Heyting algebra valued models are simpler than for 
$\{0,1\}$-valued models. For truth values algebra valued models, it is
even simpler.
We
want to prove that if $A$ is valid in all models of $\calT$ where it
has a denotation then $\calT \vdash A$.  To do so, we consider a
theory $\calT$ and we construct a model of $\calT$ such that the
formulae valid in this model are the intuitionistic theorems of $\calT$.

\begin{definition}[Lindenbaum model]
Let $\calT$ be a theory in a language $\calL$.
Let $S$ be an infinite set of constants and $\calL' = \calL \cup
S$. Let $\calM$ be the set of closed terms of $\calL'$ and $\calB_\calT$ be the
set of closed formulae of $\calL'$.
Let $\calB_\calT^{+}$ be the set of elements $A$ of $\calB_\calT$,
such that the sequent $\calT \vdash A$ is provable. 
Let $\calA = \calE$ be the set of subsets of $\calB_\calT$ 
of the form $\{(t/x)A~|~t \in \calM\}$ for some $A$. 
Notice that, in this case, the formula $A$ is unique.

The operations $\tildetop$, $\tildebot$, $\tildeimp$, $\tildeand$ and
$\tildeor$ are $\top$, $\bot$, $\Rightarrow$, $\wedge$ and $\vee$. 
The operations $\tildefa$ and $\tildeex$ are defined as follows
\begin{itemize}
\item $\tildefa~\{(t/x)A~|~ t \in \calM\} =  (\fa x~A)$, 
\item $\tildeex~\{(t/x)A~|~ t \in \calM\} =  (\ex x~A)$.
\end{itemize}
If $f$ is a function symbol, we let $\hat{f}$ be the function mapping 
$t_{1}, ..., t_{n}$ to $f(t_{1}, ..., t_{n})$. 
If $P$ is a predicate symbol, we let $\hat{P}$ be the function mapping 
$t_{1}, ..., t_{n}$ to $P(t_{1}, ..., t_{n})$. 
\end{definition}

\begin{proposition}
The algebra $\calB_\calT$ is a truth values algebra. 
\end{proposition}

\VL{\proof{The condition $1.$ to $11.$ are trivial recalling that
$\calB_\calT^{+}$ is the set of theorems of $\calT$. The condition $12.$
is a simple consequence of the definition of $\calA$ and $\calE$. 

For condition $13.$,  consider a set $A = \{(t/x)P~|~t \in \calM\}$
and $c$ a constant occurring neither in $\calT$ nor in $P$.
If all elements of $A$ are in $\calB_\calT^{+}$, $(c/x)P$
is in $\calB_\calT^{+}$ thus $\calT \vdash (c/x)P$ is provable. Thus,
$\calT \vdash \fa x~P$ is provable and $\fa x~P$ is in $\calB_\calT^{+}$, 
{\em i.e.} $\tildefa~A$ is in $\calB_\calT^{+}$.

For condition $14.$ consider an element $A$ of $\calA$,
by definition there exists a formula $P$ such that 
$A = \{(t/x)P~|~t \in \calM\}$ and we have 
$\tildefa~A = \fa x~P$ and $\tildefa~(a~\tildeimp~A) = \fa x~(a
\Rightarrow P)$. Thus, the condition rephrases 
$$\calT \vdash (\fa x~(a \Rightarrow P)) \Rightarrow a \Rightarrow \fa
x~P$$
which is obvious as the formula
$(\fa x~(a \Rightarrow P)) \Rightarrow a \Rightarrow \fa
x~P$ is intuitionisticaly provable.
The conditions $15.$, $16.$ and $17.$ are checked in a similar way.}}

\VL{
\begin{proposition}
Let $A$ be a formula and $\phi$ be an assignment mapping the free
variables of $A$ to elements of $\calM$. Notice that $\phi$ is also a
substitution and that $\phi A$ is a closed formula. Then 
$\llbracket A \rrbracket_{\phi}$ is always defined and 
$$\llbracket A \rrbracket_{\phi} = \phi A$$
\end{proposition}

\proof{By induction over the structure of $A$. We consider only the
case where $A = \fa x~B$. 
We have 
$\llbracket \fa x~B \rrbracket_{\phi} 
= \tildefa~\{\llbracket (t/x)B \rrbracket_{\phi}~|~t \in \calM\}
= \tildefa~\{\phi ((t/x)B)~|~t \in \calM\}
= \tildefa~\{(t/x)(\phi B)~|~t \in \calM\}
= \fa x~(\phi B) 
= \phi (\fa x~B)$.}

\begin{proposition}
The closed formulae valid in the Lindenbaum model of $\calT$ in
$\calL$ are intuitionistic theorems of $\calT$. 
\end{proposition}

\proof{If $A$ is valid in the Lindenbaum model then for every
assignment $\phi$, $\llbracket A \rrbracket_{\phi} \in \calB_\calT^{+}$, {\em i.e.}  $\calT
\vdash \llbracket A \rrbracket_{\phi}$ is provable.
Thus, $\calT \vdash \phi A$ is provable and in particular 
$\calT \vdash A$ is provable.}
}

\begin{proposition}[Completeness]
If $A$ is valid in all the models of $\calT$ where it is defined, then 
$\calT \vdash A$.
\end{proposition}

\VL{\proof{It is valid in the Lindenbaum model of $\calT$.}}

\medskip
\noindent
\VL{Using Proposition \ref{macneille}, we can strengthen this completeness theorem.}
\VC{Using completion, we can strengthen this completeness theorem.}

\begin{proposition}
If $A$ is valid in all the models of $\calT$ where the truth values
algebra is full, ordered and complete then $\calT \vdash A$.
\end{proposition}

\medskip
\noindent
The converse is a simple induction over proof structure.

\begin{proposition}[Soundness]
If $\calT \vdash A$ then $A$ is valid in all the models of $\calT$ where
the truth value algebra is full, ordered and complete.
\end{proposition}

\noindent
We finally get the following theorem.

\begin{theorem}
$\calT \vdash A$ if and only if 
$A$ is valid in all the models of $\calT$ where the truth values
algebra is full, ordered and complete.
\end{theorem}

\subsection{Consistency}

\begin{definition}
A theory is said to be {\em consistent} if there exists a non provable
formula in this theory.  
\end{definition}

In the completeness theorem above, we did not
assume the theory $\calT$ to be consistent. If it is not,
then the algebra of the Lindenbaum model is trivial, {\em i.e.} all
truth values are positive and every formula is valid. But we have the
following theorem.

\begin{proposition}
The theory $\calT$ is consistent if and only if it has a
$\calB$-valued model, for some non trivial full, ordered and complete truth
values algebra $\calB$.
\end{proposition}

\VL{\proof{The algebra of the Lindenbaum model and its completion are
non trivial.}} 

\section{Deduction modulo}

\subsection{Deduction modulo}

In Deduction modulo \cite{DHK,DowekWerner}, a theory is defined by a
set of axioms $\calT$ and a congruence $\equiv$ defined by a confluent
rewrite system rewriting terms to terms and atomic formulae to
formulae. The deduction rules are modified to take the congruence
$\equiv$ into account. For instance, the {\em modus ponens} rule is not
stated as usual
$$\irule{\Gamma \vdash A \Rightarrow B~~~\Gamma \vdash A}{\Gamma 
\vdash B}{}$$
but 
$$\irule{\Gamma \vdash_{\equiv} C~~~\Gamma \vdash_{\equiv} A}{\Gamma \vdash_{\equiv} B}{C \equiv
A \Rightarrow B}$$ 

In deduction modulo, there are theories for which there exists proofs
that do not normalize.
For instance, in the theory formed with the rewrite rule 
$P \lra (P \Rightarrow Q)$, the proof 
$$\hspace*{-1.25cm}
\irule{\irule{\irule{\irule{}
                             {P \vdash_{\equiv} P \Rightarrow Q}
                             {\mbox{axiom}}
                      ~~~~~~~~~~
                      \irule{}
                            {P \vdash_{\equiv} P}
                            {\mbox{axiom}}
                      }
                      {P \vdash_{\equiv} Q}
                      {\mbox{$\Rightarrow$-elim}}
               }
               {\vdash_{\equiv} P \Rightarrow Q}
               {\mbox{$\Rightarrow$-intro}}
         ~~~~~~~~~~~~~~~~~~~~~~~~~~~~~~~~~~~~~
         \irule{\irule{\irule{}
                             {P \vdash_{\equiv} P \Rightarrow Q}
                             {\mbox{axiom}}
                      ~~~~~~~~~~
                      \irule{}
                            {P \vdash_{\equiv} P}
                            {\mbox{axiom}}
                      }
                      {P \vdash_{\equiv} Q}
                      {\mbox{$\Rightarrow$-elim}}
               }
               {\vdash_{\equiv} P}
               {\mbox{$\Rightarrow$-intro}}
       }
       {\vdash_{\equiv} Q}
       {\mbox{$\Rightarrow$-elim}}$$
does not normalize and, moreover, the formula $Q$ has no normal
proof. 

But, as we shall see, in some other theories, such as the theory
formed with the rewrite rule $P \lra (Q \Rightarrow P)$, all proofs
strongly normalize.

In deduction modulo, like in predicate logic, normal proofs of a
sequent of the form $\vdash_{\equiv} A$ always end with an introduction rule.
Thus, when a theory can be expressed in deduction modulo with rewrite
rules only, {\em i.e.} with no axioms, in such a way that proofs
modulo these rewrite rules strongly normalize, then the theory is consistent,
it has the disjunction property and the witness property, various
proof search methods for this theory are complete, ...

Many theories can be expressed this way in deduction modulo, in
particular arithmetic \cite{Peano} and simple type theory \cite{DHK2} 
and the notion of cut
of deduction modulo subsumes the {\em ad hoc} notions of cut defined
for these theories.

\subsection{Models}

\begin{definition}[Model]
Let $\calT, \equiv$ be a theory in deduction modulo.
The $\calB$-valued structure $\calM$ is said to be {\em a model of} the theory
$\calT, \equiv$ if all axioms of $\calT$ are valid in $\calM$ and for
all terms or formulae $A$ and $B$ such that 
$A \equiv B$ and assignments $\phi$, $\llbracket A
\rrbracket_\phi$ and $\llbracket B \rrbracket_\phi$ are defined and
$\llbracket A \rrbracket_\phi = \llbracket B \rrbracket_\phi$.
\end{definition}

\begin{example}
Let $\calB$ be an arbitrary truth value algebra, then the theory $P
\lra (Q \Rightarrow R)$ has a $\calB$-valued model. We take $\hat{P} =
(\tildetop~\tildeimp~\tildetop)$ and $\hat{Q} = \hat{R} = \tildetop$.
\end{example}

\begin{example}
Let $\calB$ be an arbitrary full, ordered and complete truth value
algebra, then the theory $P \lra (Q \Rightarrow P)$ has a
$\calB$-valued model.  The function $a \mapsto (\tildebot~\tildeimp~a)$ is
monotonous for the order $\sqsubseteq$ and this order is complete.
Hence, it has a fixed point $b$.  We define a $\calB$-valued model of this
theory by
$\hat{P} = b$ and $\hat{Q} = \tildebot$. 

In the same way, if $\calB$ be an arbitrary full, ordered and complete
truth value algebra, then the theory $P \lra (\bot \Rightarrow P)$ has
a $\calB$-valued model.
\end{example}

\begin{example}
The theory $P \lra (P \Rightarrow Q)$ has a $\{0,1\}$-valued model
($\hat{P} = \hat{Q} = 1$), but no $\calT_1$-valued model.
Indeed there is no $0$ in the line $0$ of the table of the function
$\tildeimp$ of $\calT_1$, no $I$ in the line $I$ and no $1$ in
the line $1$.
\end{example}

\subsection{Soundness and completeness}

To extend the completeness and the soundness theorem to deduction
modulo, we replace terms by classes of congruent terms and formulae by
classes of congruent formulae.

\begin{definition}
Let $\calT, \equiv$ be a theory in a language $\calL$.
Let $S$ be an infinite set of constants and $\calL' = \calL \cup S$. 
We define an equivalence relation $\sim$ on formulae of ${\cal L}'$
inductively as the smallest congruence such that 
if $A \equiv B$ then $A \sim B$,
if $A \sim B$ and $A' \sim B'$ then 
$(A \wedge A') \sim (B \wedge B')$, 
$(A \vee A') \sim (B \vee B')$, and
$(A \Rightarrow A') \sim (B \Rightarrow B')$, 
and if for each term $t$ there exist a term $u$ such that 
$(t/x)A \sim (u/x)B$ 
and for each term $u$ there exist a term $t$ such that 
$(u/x)B \sim (t/x)A$ 
then $\fa x~A \sim \fa x~B$ and 
$\ex x~A \sim \ex x~B$. 
\end{definition} 

Remark that if we consider the congruence $\equiv$ defined by the rewrite rule
$f(f(x)) \lra x$ we have $\fa x~P(x) \not\equiv \fa x~P(f(x))$ but we have
$\fa x~P(x) \sim \fa x~P(f(x))$ as 
$P(x)$ and $P(f(x))$ have the same
instances (the instance $t$ in one formula corresponds to the
instance $f(t)$ in the other).  

\begin{proposition}
If $t \equiv u$ and $A \sim B$ then $(t/x)A \sim (u/x)B$.
\end{proposition}

\proof{By induction on the derivation of $A \sim B$.}

\begin{proposition}
\label{miraculous}
If $A \sim B$ then $A \Leftrightarrow B$ is provable modulo $\equiv$.
\end{proposition}

\proof{By induction on the derivation of $A \sim B$ if $A = \fa x~A'$
and $B = \fa x~B'$ and 
for each term $t$ there exist a term $u$ such that 
$(t/x)A \sim (u/x)B$ 
and for each term $u$ there exist a term $t$ such that 
$(u/x)B \sim (t/x)A$ 
then let $c$ be a constant of $S$ occurring neither in $A$ nor
in $B$. We have to prove the sequent $\fa x~A' \vdash (c/x)B'$. 
Let $t$ be a term such that $(t/x)A' \sim (c/x)B'$, 
by induction hypothesis we get $(t/x)A' \Leftrightarrow (c/x)B'$, and
as we have 
$\fa x~A'$ we can deduce $(t/x)A'$, 
and thus $(c/x)B'$.}

\begin{definition}[The Lindenbaum model]
Let $\calT, \equiv$ be a theory in a language $\calL$.
Let $S$ be an infinite set of constants and $\calL' = \calL \cup S$. 
Let $\calM$ be the set of $\equiv$-classes of 
closed terms of $\calL'$ and $\calB$ be the
set of $\sim$-classes of 
closed formulae of $\calL'$.
Let $\calB^{+}$ be the set of elements $A$ of $\calB$,
such that the sequent $\calT \vdash_{\equiv} A$ is provable. 
Let $\calA = \calE$ be the set of subsets of $\calB$ 
of the form $\{(t/x)A / \sim~|~t \in \calM\}$ for some $A$. 

The operations $\tildetop$, $\tildebot$, $\tildeimp$, $\tildeand$ and
$\tildeor$ are $\top$, $\bot$, $\Rightarrow$, $\wedge$ and $\vee$
extended to $\sim$-classes.
To define the operations $\tildefa$ and $\tildeex$, we choose for each
element $a$ of $\calA$ and $\calE$ a formula $A$ such that $a =
\{(t/x)A/\sim~|~ t \in \calM\}$ and we let 
\begin{itemize}
\item $\tildefa~a =  (\fa x~A)/\sim$, 
\item $\tildeex~a =  (\ex x~A)/\sim$.
\end{itemize}
Notice that the elements $(\fa x~A)/\sim$ and $(\ex x~A)/\sim$ 
are independent of the choice of $A$. 

If $f$ is a function symbol, we let $\hat{f}$ be the function mapping
the classes of 
$t_{1}, ..., t_{n}$ to that of $f(t_{1}, ..., t_{n})$. 
If $P$ is a predicate symbol, we let $\hat{P}$ be the function mapping 
the classes of $t_{1}, ..., t_{n}$ to that of $P(t_{1}, ..., t_{n})$. 
\end{definition}

\begin{proposition}
Let $A$ be a formula and $\phi$ be a assignment mapping the free
variables of $A$ to elements of $\calM$. Notice that $\phi$ is also a
substitution and that $\phi A$ is a closed formula. Then 
$\llbracket A \rrbracket_{\phi} = \phi A / \sim$.
\end{proposition}

\proof{By induction over the structure of $A$. We consider only the
case where $A = \fa x~B$. 
We have 
$\llbracket \fa x~B \rrbracket_{\phi} 
= \tildefa~\{\llbracket B \rrbracket_{\phi + (x = a)}~|~a \in \calM\}$.
By induction hypothesis, 
$\llbracket \fa x~B \rrbracket_{\phi} 
= \tildefa~\{(a/x)\phi B / \sim |~a \in \calM\}
= (\fa x~\phi B) / \sim 
= \phi (\fa x~B) / \sim$.}

\begin{proposition}
The algebra $\calB$ is a truth values algebra. 
\end{proposition}

\VL{
\proof{The condition $1.$ to $11.$ are trivial using the fact that
$\calB^{+}$ is the set of theorems of $\calT, \equiv$. The condition $12.$
is a simple consequence of the definition of $\calA$ and $\calE$. 

For condition $13.$,  consider a set $A$ in $\calA$ and 
$P$ the formula associated to this set, 
and $c$ a constant occurring neither in $\calT$
nor in the rewrite system defining the congruence $\equiv$,
nor in $P$.
If all elements of $A$ are in $\calB^{+}$, $(c/x)P$
is in $\calB^{+}$ thus $\calT \vdash_{\equiv} (c/x)P$ is
provable. Thus, $\calT \vdash_{\equiv} \fa x~P$ is provable and $\fa
x~P$ is in $\calB^{+}$, {\em i.e.} $\tildefa~A$ is in $\calB^{+}$.

For condition $14.$, consider an element $A$ of $\calA$ and the formula
$P$ associated to this element and $Q$ the formula associated to 
the set $a~\tildeimp~A$. By Proposition \ref{miraculous}, the formula 
$(\fa x~(a \Rightarrow P)) \Leftrightarrow (\fa x~Q)$ is provable. 
We have 
$\tildefa~A = \fa x~P$ and $\tildefa~(a~\tildeimp~A) = \fa x~Q$. 
Thus, the condition rephrases 
$$\calT \vdash_{\equiv} \fa x~Q \Rightarrow a \Rightarrow \fa x~P$$
which is a consequence of the fact that the formula
$(\fa x~(a \Rightarrow P)) \Leftrightarrow (\fa x~Q)$ is provable. 
The conditions $15.$, $16.$ and $17.$ are checked in a similar way.}
}

\VL{
\begin{proposition}
The closed formulae valid in the Lindenbaum model of $\calT, \equiv$ in
$\calL$ are intuitionistic theorems of $\calT, \equiv$. 
\end{proposition}

\proof{If $A$ is valid in the Lindenbaum model, then for every
assignment $\phi$, $\llbracket A \rrbracket_{\phi} \in \calB^{+}$,
{\em i.e.}  
$\calT
\vdash_{\equiv} \llbracket A \rrbracket_{\phi}$ is provable in
deduction modulo.  Thus, 
$\calT \vdash_{\equiv} \phi A$ is provable in deduction modulo and in
particular $\calT \vdash_{\equiv} A$ is provable in deduction modulo.}
}

\begin{proposition}[Completeness]
If $A$ is valid in all the models of $\calT,\equiv$ where it is defined, then 
$\calT \vdash_{\equiv} A$.
\end{proposition}

\VL{\proof{It is valid in the Lindenbaum model of $\calT,\equiv$.}}

\medskip
\noindent
\VL{Using Proposition \ref{macneille}, we can strengthen this completeness theorem.}
\VC{Using completion, we can strengthen this completeness theorem.}

\begin{proposition}
If $A$ is valid in all the models of $\calT, \equiv$ where the truth values
algebra is full, ordered and complete then $\calT \vdash_{\equiv} A$.
\end{proposition}

\medskip
\noindent
The converse is a simple induction over proof structure.

\begin{proposition}[Soundness]
If $\calT \vdash_{\equiv} A$ then $A$ is valid in all the models of
$\calT, \equiv$ where the truth value algebra is full, ordered and
complete.
\end{proposition}

\noindent
We finally get the following theorem.

\begin{theorem}
$\calT \vdash_{\equiv} A$ if and only if 
$A$ is valid in all the models of $\calT,\equiv$ where the truth values
algebra is full, ordered and complete.
\end{theorem}

\subsection{Consistency}

\begin{proposition}
\label{link}
The theory $\calT, \equiv$ is consistent if and only if it has a
$\calB$-valued model, for some non trivial full, ordered and complete truth
values algebra $\calB$.
\end{proposition}

\VL{\proof{The algebra of the Lindenbaum model and its completion are
non trivial.}} 

\section{Super-consistency}

\subsection{Definition}

By Proposition \ref{link}, a theory is consistent if it has a
$\calB$-valued model for some non trivial full, ordered and complete truth
values algebra. We now strengthen this condition and require that
the theory has a $\calB$-valued model for all full, ordered and complete
truth values algebras $\calB$.

\begin{definition}[Super-consistent]
A theory $\calT, \equiv$ in deduction modulo is
{\em super-consistent} if it has a $\calB$-valued model for all full, ordered
and complete truth values algebras $\calB$. 
\end{definition}

Notice that, as there exists non trivial full, ordered and complete truth
values algebras ({\em e.g.} $\{0,1\}$), super-consistent theories are
consistent.

\subsection{Examples of super-consistent theories}

We have seen that the theories $P \lra (Q \Rightarrow R)$ and $P \lra
(Q \Rightarrow P)$ are super-consistent, but that the theory $P \lra
(P \Rightarrow Q)$ is not. We give other examples of super-consistent
theory. In particular, we show that all the theories that have been
proved to have the strong normalization property in
\cite{DowekWerner,Peano} 
\VC{{\em i.e.} arithmetic, simple type theory, the theories defined by a 
confluent, terminating and quantifier free rewrite system, the 
theories defined by a 
confluent, terminating and positive rewrite systems 
and 
the theories defined by a positive rewrite systems such that
each atomic formula has at most one one-step reduct}
are super-consistent. \VC{In this abstract, we detail only the
case of simple type theory.}

\begin{definition}[Simple type theory]
Simple type theory is a many-sorted theory defined
as follows. The sorts are inductively defined by 
$\iota$ and $o$ are sorts and if $T$ and $U$ are sorts then $T \ra U$
is a sort.
The language contains the constants
$S_{T,U,V}$ of sort $(T \ra U \ra V) \ra (T \ra U) \ra T \ra V$,
$K_{T,U}$ of sort $T \ra U \ra T$, 
$\dot{\top}$ of sort $o$ and $\dot{\bot}$ of sort $o$,
$\dot{\Rightarrow}$, $\dot{\wedge}$ and $\dot{\vee}$ 
of sort $o \ra o \ra o$, 
$\dot{\fa}_{T}$ and $\dot{\ex}_T$ of sort $(T \ra o) \ra o$,
the function symbols
$\alpha_{T,U}$ of rank $\langle T \ra U, T, U \rangle$ 
and the predicate symbol
$\varepsilon$ of rank $\langle o \rangle$.
The rules are
\begin{eqnarray*}
\alpha(\alpha(\alpha(S_{T,U,V},x),y),z) &\lra& \alpha(\alpha(x,z),\alpha(y,z))\\
\alpha(\alpha(K_{T,U},x),y) &\lra& x\\
\varepsilon(\dot{\top}) &\lra& \top\\
\varepsilon(\dot{\bot}) &\lra& \bot\\
\varepsilon(\alpha(\alpha(\dot{\Rightarrow},x),y)) &\lra&
\varepsilon(x) \Rightarrow \varepsilon(y)\\ 
\varepsilon(\alpha(\alpha(\dot{\wedge},x),y)) &\lra&
\varepsilon(x) \wedge \varepsilon(y)\\ 
\varepsilon(\alpha(\alpha(\dot{\vee},x),y)) &\lra&
\varepsilon(x) \vee \varepsilon(y)\\ 
\varepsilon(\alpha(\dot{\fa}_T,x)) &\lra& \fa y~\varepsilon(\alpha(x,y))\\
\varepsilon(\alpha(\dot{\ex}_T,x)) &\lra& \ex y~\varepsilon(\alpha(x,y))\\
\end{eqnarray*}
\end{definition}

\begin{proposition}
\label{stt}
Simple type theory is super-consistent.
\end{proposition}

\proof{Let $\calB$ be a full truth values algebra. The model 
$\calM_{\iota} = \{0\}$, $\calM_{o} = {\calB}$, 
$\calM_{T \ra U} = \calM_{U}^{\calM_{T}}$,
$\hat{S}_{T,U,V} = a \mapsto (b \mapsto (c \mapsto a(c)(b(c))))$,
$\hat{K}_{T,U} = a \mapsto (b \mapsto a)$,
$\hat{\alpha}(a,b) = a(b)$,
$\hat{\varepsilon}(a) = a$,
$\hat{\dot{\top}} = \tildetop$, 
$\hat{\dot{\bot}} =\tildebot$,
$\hat{\dot{\Rightarrow}} = \tildeimp$, 
$\hat{\dot{\wedge}} = \tildeand$,
$\hat{\dot{\vee}} = \tildeor$,
$\hat{\dot{\fa}}_T = a \mapsto \tildefa (Range(a))$,
$\hat{\dot{\ex}}_T = a \mapsto \tildeex (Range(a))$
where $Range(a)$ is the range of the function $a$, is a $\calB$-valued model
of simple type theory.}

\VL {
\begin{definition}[Arithmetic]
Arithmetic is a many-sorted theory defined
as follows. The sorts are $\iota$ and $\kappa$.
The language contains 
the constant $0$ of sort $\iota$, the function symbols $S$ and $\pred$ of
rank $\langle \iota, \iota \rangle$ and $+$ and $\times$ of rank  
$\langle \iota, \iota, \iota \rangle$,
the predicate symbols $=$ of rank $\langle \iota, \iota \rangle$,
$\nulll$ and $N$ of rank 
$\langle \iota \rangle$
and $\in$ of rank $\langle \iota, \kappa \rangle$ 
and for each formula $P$ in the language 
$0$, $S$, $\pred$, $+$, $\times$, $=$, $\nulll$ and $N$ and 
whose free variables are among $x$, $y_{1}, \dots, y_n$ of sort
$\iota$, the function symbol $f_{x, y_1, ..., y_n, P}$ of rank 
$\langle \iota, \dots, \iota, \kappa \rangle$. 
The rewrite rules are
$$x \in f_{x, y_1, ..., y_n, P}(y_{1}, \dots, y_n) \lra P$$
$$y = z \lra \fa p~(y \in p \Rightarrow z \in p)$$
$$N(n) \lra   \fa p~(0 \in p \Rightarrow \fa
  y~(N(y) \Rightarrow y \in p \Rightarrow S(y) \in p) \Rightarrow n \in p)$$
$$\pred(0) \lra 0$$
$$\pred(S(x)) \lra x$$
$$\nulll(0) \lra \top$$
$$\nulll(S(x)) \lra \bot$$
$$0 + y \lra y$$
$$S(x) + y \lra S(x+y)$$
$$0 \times y \lra 0$$
$$S(x) \times y \lra x \times y + y$$
\end{definition}

\begin{proposition}
Arithmetic is super-consistent.
\end{proposition}

\proof{Let $\calB$ be a full, ordered and complete truth value algebra.
We take $\calM_{\iota} = {\mathbb N}$, $\calM_{\kappa} = {\calB}^{\mathbb N}$. 
The denotations of $0$, $S$, $+$, $\times$, $\pred$ are 
obvious. We take $\hat{\nulll} (0) = \tildetop$, 
$\hat{\nulll} (n) = \tildebot$ if $n \neq 0$.
The denotation of $\in$ is the function mapping $n$ and $f$ to $f(n)$. 
Then, we can define the denotation of 
$\fa p~(y \in p \Rightarrow z \in p)$
and the denotation of $=$ accordingly.

To define the denotation of $N$, for each function $f$ of
${\calB}^{\mathbb N}$ we can define an 
interpretation $\calM_{f}$ of the language of the formula
$$\fa p~(0 \in p \Rightarrow \fa y~(N(y) \Rightarrow y \in p
\Rightarrow S(y) \in p) \Rightarrow n \in p)$$
where the symbol $N$ is interpreted by the function $f$.
We define the function 
$\Phi$ from ${\calB}^{\mathbb N}$ to  ${\calB}^{\mathbb N}$ mapping $f$ to
the function mapping the natural number $x$ to 
the truth value
$$\llbracket \fa p~(0 \in p \Rightarrow \fa y~(N(y) \Rightarrow y \in p
\Rightarrow S(y) \in p) \Rightarrow n \in p)\rrbracket_{x/n}^{\calM_{f}}$$

The order on ${\calB}^{\mathbb N}$ defined by $f \sqsubseteq g$ if for all
$n$,
$f(n) \sqsubseteq g(n)$ is a complete order and the function $\Phi$ is 
monotonous as the occurrence of $N$ is positive in 
$$\fa p~(0 \in p \Rightarrow \fa y~(N(y) \Rightarrow y \in p
\Rightarrow S(y) \in p) \Rightarrow n \in p)$$ Hence it has a fixed
point $g$. We interpret the symbol $N$ by the function $g$.  Finally,
the denotation of the symbols of the form $f_{x, y_1, ..., y_n, P}$ is
defined in the obvious way.}
}

\VL{
\begin{proposition}[Quantifier free]
A theory defined by a confluent and terminating rewrite systems such that
no quantifier appears in the rewrite rules is super-consistent. 
For instance, the theory defined by the rewrite system 
$P \lra Q \Rightarrow R$ is super-consistent.
\end{proposition}

\proof{
Let $\calB$ be an arbitrary full truth value algebra.
We associate an element of $\calB$ to each normal closed
quantifier free formula as follows: 
if $A$ is atomic then $|A| = \tildetop$, 
$|\top| = \tildetop$,
$|\bot| = \tildebot$,
$|A \Rightarrow B| = |A|~\tildeimp~|B|$, 
$|A \wedge B| = |A|~\tildeand~|B|$, 
$|A \vee B| = |A|~\tildeor~|B|$. 
We then define a $\calB$-valued model as follows: 
$\calM$ is the set of normal closed terms, 
$\hat{f}(t_{1}, \dots, t_{n}) = f(t_{1}, \dots, t_{n})\downarrow$, 
$\hat{P}(t_{1}, \dots, t_{n}) = |P(t_{1}, \dots, t_{n})\downarrow|$
where $a \downarrow$ is the normal form of the $a$.
}
}
\VL{
\begin{proposition}[Positive terminating]
\label{positive1}
A theory defined by a confluent and terminating rewrite systems such that
all atomic formulae appear at positive occurrences in the rewrite rules
is super-consistent.
For instance the theory defined by the rewrite system 
$P(0) \lra \fa x~P(x)$ is super-consistent.
\end{proposition}
}
\VL{
\proof{Consider a full, ordered and complete truth value algebra
${\cal B}$. Let $\calT$ be the set of closed terms and $\calM = \calT
/ \equiv$.  Let $\hat{f}$ be the function mapping the classes $e_{1},
..., e_{n}$ to the class of the term $f(t_{1}, \dots, t_{n})$ where
$t_{1}, ..., t_{n}$ are elements of $e_{1}, ..., e_{n}$ (since the
relation $\equiv$ is a congruence, this class does not depend of the
choice of representatives).
Let $\calC$ be the set of models that have the domain $\calM$, the truth
value algebra $\calB$ and where the function symbols $f$ are
interpreted by the functions $\hat{f}$. Two models of $\calC$
differ only by the interpretation of predicate symbols. 
Let $\calM_1$ and $\calM_2$ be two models of the class
$\calC$, we say that $\calM_1 \sqsubseteq \calM_2$ if and only
if for every predicate symbol $P$ and sequence of elements of $\calM$ 
$e_{1},
\dots, e_{n}$, we have
$$\hat{P}^{\calM_1}(e_{1}, \dots, e_{n}) \sqsubseteq
\hat{P}^{\calM_2}(e_{1}, \dots, e_{n})$$ 
As the algebra $\calB$ is complete, the set $\calC$ is a complete
lattice for the order $\sqsubseteq$.
Let $\calF$ be the function from $\calC$ to $\calC$ 
defined by
$$\hat{P}^{\calF(\calM)}(e_{1}, \dots, e_{n}) = \llbracket P(e_{1}, \dots,
e_{n}) \downarrow \rrbracket^{\calM}_{\emptyset}$$ 
As all atomic formulae appear at positive occurrences in the rewrite
system, the function $\calF$ is monotone.  Hence, as $\calC$ is 
a complete lattice, it has a fixed point. This fixed point is a
$\calB$-valued model of the theory.}}

\VL{\begin{proposition}[Positive deterministic]
A theory defined by a rewrite systems such that
each atomic formula has at most one one-step reduct and 
all atomic formulae appear at positive occurrences in the rewrite rules
is super-consistent. For instance, the theory defined by the rewrite system 
$P \lra (P \wedge P)$ is super-consistent.
\end{proposition}

\proof{
As in the proof of Proposition \ref{positive1}, 
we consider a full, ordered and complete truth value algebra ${\cal
B}$ and we define the complete lattice $\calC$ of models.

Let $\calF$ be the function from $\calC$ to $\calC$ 
defined by
$$\hat{P}^{\calF(\calM)}(t_{1}, \dots, t_{n}) = \llbracket P(t_{1}, \dots,
t_{n}) +\rrbracket^{\calM}_{\emptyset}$$
where $A+$ is the unique one-step reduct of $A$ if it exists and $A$
otherwise. 
As all atomic formulae appear at positive occurrences in the rewrite
system, the function $\calF$ is monotone.  Hence, as $\calC$ is 
a complete lattice, it has a fixed point. This fixed point is a
$\calB$-valued model of the theory.}
}

\subsection{Normalization}

We have seen that the theory $P \lra (P \Rightarrow Q)$, that does not
have the strong normalization property, is consistent but not
super-consistent, {\em i.e.} it has $\calB$-valued models for some non
trivial, full, ordered and complete truth values algebras $\calB$, but
not all.
We prove now that, in contrast, all super-consistent theories have the
strong normalization property.  To prove this, we build a particular full,
ordered and complete truth values algebra: the algebra of reducibility
candidates.

We refer, for instance, to \cite{DowekWerner} for the definition of
proof-terms, neutral proof-terms and of proof-term reduction $\rhd$
and we define the following operations on sets of proofs.

\begin{definition}\label{operations}

\begin{itemize}
\item The set $\tildetop$ is the set of strongly normalizing
proof-terms. 

\item The set $\tildebot$ is the set of strongly normalizing
proof-terms.

\item If $a$ and $b$ are two sets of proofs-terms, then 
$a~\tildeimp~b$ is the set of strongly normalizing proof-terms
$\pi$ such that if $\pi$ reduces to 
$\lambda \alpha~\pi_{1}$ then for every $\pi'$ in $a$, $(\pi' /
\alpha)\pi_{1}$ is in $b$. 

\item If $a$ and $b$ are two sets of proof-terms, then 
then $a~\tildeand~b$ is the set of strongly normalizing proof-terms
$\pi$ such that if $\pi$ reduces to
$\langle \pi_{1},\pi_{2} \rangle$ then $\pi_{1}$ is in $a$ and 
$\pi_{2}$ is in $b$.

\item 
If $a$ and $b$ are two sets of proof-terms, then 
$a~\tildeor~b$
is the set of strongly normalizing proof-terms $\pi$ such that 
if $\pi$ reduces to
$i(\pi_{1})$ (resp. $j(\pi_{2})$) then $\pi_{1}$ (resp. $\pi_{2}$) is
in $a$ (resp. $b$). 

\item 
If $A$ is a set of sets of proof-terms, then 
$\tildefa~A$ is
the set of strongly normalizing proof-terms $\pi$ such that 
if $\pi$ reduces to 
$\lambda x~\pi_{1}$ then for every term $t$ 
and every element $a$ of $A$, $(t/x)\pi_{1}$ is in $a$.

\item 
If $A$ is a set of sets of proof-terms, then 
$\tildeex~A$ is the set of strongly normalizing proof-terms
$\pi$ such that if $\pi$ reduces to
$\langle t,\pi_{1} \rangle$, there exists an element $a$ of 
$A$ such that $\pi_{1}$ is an element of $a$.
\end{itemize}
\end{definition}

\begin{definition}[Reducibility candidate]
A set $R$ of proof-terms is a {\em reducibility candidate} if 
\begin{itemize}
\item if $\pi \in R$, then $\pi$ is strongly normalizable,
\item if $\pi \in R$ and $\pi \rhd^{*} \pi'$ then $\pi' \in R$,
\item if $\pi$ is neutral and 
if for every $\pi'$ such that $\pi \rhd^{1} \pi'$, $\pi' \in R$ then 
$\pi \in R$. 
\end{itemize}
\end{definition}

\begin{proposition} 
The set of reducibility candidates is closed by the operations of 
Definition \ref{operations}. 
\end{proposition}

\VL{\proof{
All the cases are similar. Let us detail, for instance, the case of
the operation $\tildefa$. Let $A$ be a set of candidates and let us
prove that the set $\tildefa~A$ is a candidate.

By definition, all proof-terms of $\tildefa~A$ are strongly
normalizing.  Closure by reduction is a simple consequence of the fact
that if $\pi \in \tildefa~A$ and $\pi \rhd^{*} \pi'$ then $\pi$ is
strongly normalizing and thus so is $\pi'$ and if $\pi'$ reduces to
$\lambda x~\pi_1$ then so does $\pi$ and thus for every term $t$ and
every $a$ in $A$, $(t/x)\pi_1$ is in $a$.

Now, assume that $\pi$ is a neutral proof-term and that for every $\pi'$
such that $\pi \rhd^1 \pi'$, $\pi' \in \tildefa~A$. We want to prove
that $\pi$ is in $\tildefa~A$. Following the definition of
$\tildefa~A$, we first prove that $\pi$ is strongly normalizing and
then that if it reduces to $\lambda x~\pi_1$ then for every term $t$
and every $a$ in $A$, $(t/x)\pi_1$ is in $a$.

Consider a reduction sequence issued from $\pi$.  If it is empty it is
finite. Otherwise it has the form $\pi \rhd^1 \pi_{2} \rhd^1 ...$ The
proof-term $\pi_{2}$ is an element of $\tildefa~A$ thus it is strongly
normalizing and the reduction sequence is finite.  If $\pi$ reduces to
$\lambda x~\pi_1$ then consider a reduction sequence from $\pi$ to
$\lambda x~\pi_1$. As $\pi$ is neutral and $\lambda x~\pi_1$ is not,
this sequence is not empty.  Thus, there exists a proof-term
$\pi_2$ such that $\pi \rhd^1 \pi_{2} \rhd^{*} \lambda x~\pi_1$.  We have
$\pi_{2} \in \tildefa~A$ and thus for every term $t$ and every $a$ in
$A$, $(t/x)\pi_1$ is in $a$.}
}

\begin{definition}[The algebra of reducibility candidates]
\label{acr}
The set $\calB$ is the set of reducibility candidates. 
The set $\calB^{+}$ may be any set closed by intuitionistic deduction
rules, {\em e.g.} the set of all candidates.
The sets $\calA$ and $\calE$ are $\wp(\calB)$. 
The operations are those of definition \ref{operations}.
The order $\sqsubseteq$ is inclusion. 
\end{definition}

\begin{theorem}[Normalization]
If the theory $\calT, \equiv$ is super-consistent, then
all proofs strongly normalize in $\calT, \equiv$.
\end{theorem}

\proof{Consider the full, ordered and complete truth values algebra
$\calB$ of reducibility candidates. As it is super-consistent, the
theory $\calT, \equiv$ has a $\calB$-valued model. This model is a
reducibility candidate valued model of $\equiv$ \cite{DowekWerner},
called {\em pre-models} there. Hence all proofs strongly normalize in
$\calT, \equiv$.}

\medskip
An alternative would be to define the set of candidates directly as
the smallest set of sets of proofs closed by the operations of
definition \ref{operations} and arbitrary intersections, like
\cite{Parigot}.

\medskip
Notice that the pre-order $\leq$ is trivial and thus not
antisymmetric. Hence, the truth values algebra of reducibility
candidates is not a Heyting algebra.
The fact that the choice of the set $\calB^{+}$ is immaterial is due
to the fact that $\calB^{+}$ matters for the interpretation of axioms 
but not for that of the congruence and cut elimination is a property
of the congruence of a theory, not of its axioms.

\section{Conclusion}

We have generalized the notion of Heyting algebra into a notion of
{\em truth values algebra} and proved that a theory is consistent if
and only if it has a $\calB$-valued model for some non trivial full, ordered
and complete truth values algebra $\calB$.  Unlike Heyting algebra
valued models, truth values algebra valued models allow to distinguish
computational equivalence from provable equivalence.

When a theory has a $\calB$-valued model for all full, ordered and complete
truth values algebras, it is said to be super-consistent and all
proofs strongly normalize in this theory.  Proving strong
normalization by proving super-consistency is easier than proving
strong normalization directly. For instance the proof that simple type
theory is super-consistent (Proposition \ref{stt}) takes only a few
lines.  All the technicalities related to the notion of reducibility
candidate are now hidden in the proof that super-consistency implies
strong normalization and are not used in the proof that the theory of
interest is super-consistent.

The notion of super-consistency is a model theoretic sufficient
condition for strong normalization. It remains to understand if it
also a necessary condition or if some theories have the strong
normalization property without being super-consistent. To prove that
strong normalization implies super-consistency, we might need to
restrict further the notion of super-consistency. For instance, we
have already restricted it by considering only ordered and complete
truth values algebras. Indeed, without such a completeness property,
we could not use the fixed point theorem to prove that the theory $P
\lra (\bot \Rightarrow P)$ had a ${\cal B}$-valued model for all $\calB$, and
indeed, this theory does not have a $\calT_2$-valued model.  Thus, the fact
that the algebra of reducibility candidates, ordered by inclusion, is
complete seems to be an essential property that needs to be kept when
abstracting on reducibility candidates. It remains to understand if
there are other essential properties of candidates that need to be
kept this way, so that strong normalization may imply
super-consistency.

\section{Acknowledgments}

Thierry Coquand suggested to characterize truth values algebras as
pseudo-Heyting algebras.  Lisa Allali, Fr\'ed\'eric Blanqui, Olivier
Hermant and Milly Maietti provided many helpful comments on a previous
version of the paper.


\begin{thebibliography}{99.}
\bibitem{Andrews} 
P.B.~Andrews.  Resolution in type theory.  {\em The Journal of
Symbolic Logic}, 36(3):414--432, 1971.

\bibitem{DeMarcoLipton} 
M.~De~Marco and J.~Lipton.  Completeness and cut-elimination in the
intuitionistic theory of types.  {\em Journal of Logic and
Computation}, 15:821--854, 2005.

\bibitem{DHK} 
G.~Dowek, T.~Hardin, and C.~Kirchner.  Theorem proving modulo.  {\em
Journal of Automated Reasoning}, 31:32--72, 2003.

\bibitem{DHK2} 
G.~Dowek, T.~Hardin, and C.~Kirchner.  
HOL-lambda-sigma: an intentional first-order expression of
higher-order logic.
{\em Mathematical Structures in Computer Science}, 11:1--25, 2001.

\bibitem{DowekWerner}
G.~Dowek and B.~Werner.  Proof normalization modulo.  {\em The Journal
of Symbolic Logic}, 68(4):1289--1316, 2003.

\bibitem{Peano}
G.~Dowek and B.~Werner. Arithmetic as a theory modulo. 
J. Giesel (Ed.), {\em Term rewriting and applications}, 
Lecture Notes in Computer Science 3467, Springer-Verlag, 2005, pp. 423-437. 

\bibitem{Girard}
J.-Y. Girard.  Une extension de l'interpr\'etation de {G}\"odel \`a
l'analyse, et son application \`a l'\'elimination des coupures dans
l'analyse et la th\'eorie des types.  In J.~Fenstad, editor, {\em
2$^{\mbox{nd}}$ Scandinavian Logic Symposium}, pages 63--92. North
Holland, 1971.

\bibitem{Hermant2003}
O.~Hermant. A model based cut elimination proof.  In {\em
2$^{\mbox{nd}}$ St-Petersbourg Days in Logic and Computability},
2003.

\bibitem{HermantThese}
O.~Hermant.  {\em M\'ethodes s\'emantiques en d\'eduction modulo}.
Doctoral Thesis. Universit\'e de Paris 7, 2005.

\bibitem{Hermant2005}
O.~Hermant.  Semantic cut elimination in the intuitionistic sequent
calculus.  In P.~Urzyczyn, editor, {\em Typed Lambda Calculi and
Applications}, number 3461 in Lectures Notes in Computer Science,
pages 221--233, 2005.

\bibitem{Okada}
M.~Okada.  A uniform semantic proof for cut elimination and
completeness of various first and higher order logics.  {\em
Theoretical Computer Science}, 281:471--498, 2002.

\bibitem{Parigot}
M.~Parigot.
{\em Strong normalization for the second orclassical natural
  deduction.}
Logic in Computer Science, 39--46, 1993.

\bibitem{Prawitz} 
D.~Prawitz. Hauptsatz for higher order logic. {\em The Journal of
Symbolic Logic}, 33:452--457, 1968.

\bibitem{Rasiowa}
H.~Rasiowa and R.~Sikorski. {\em The mathematics of metamathematics}.
Polish Scientific Publishers, 1963.

\bibitem{Tait66}

W.~W. Tait.  A non constructive proof of {G}entzen's {H}auptsatz for
second order predicate logic.  {\em Bulletin of the American
Mathematical Society}, 72:980--983, 1966.

\bibitem{Tait67} W.~W. Tait. Intentional interpretations of
functionals of finite type {I}.  {\em The Journal of Symbolic Logic},
32:198--212, 1967.

\bibitem{Takahashi}
M.~o. Takahashi. A proof of cut-elimination theorem in simple type theory.
{\em Journal of the Mathematical Society of Japan}, 19:399--410, 1967.
\end{thebibliography}
\end{document}